\documentclass[aps,prb,twocolumn,byrevtex]{revtex4}

\usepackage{times,mathptm}
\usepackage{graphicx}
\usepackage[dvips]{epsfig}
\usepackage{multirow}

\begin{document}

\title{Competing charge density wave and antiferromagnetism of metallic atom wires in GaN(10${\overline{1}}$0) and ZnO(10${\overline{1}}$0)}
\author{Yoon-Gu Kang, Sun-Woo Kim, and Jun-Hyung Cho$^{*}$}
\affiliation{Department of Physics and Research Institute for Natural Sciences, Hanyang University, 222 Wangsimni-ro, Seongdong-gu, Seoul, 133-791, Korea}

\date{\today}

\begin{abstract}
Low-dimensional electron systems often show a delicate interplay between electron-phonon and electron-electron interactions, giving rise to interesting quantum phases such as the charge density wave (CDW) and magnetism. Using the density-functional theory (DFT) calculations with the semilocal and hybrid exchange-correlation functionals as well as the exact-exchange plus correlation in the random-phase approximation (EX + cRPA), we systematically investigate the ground state of the metallic atom wires containing dangling-bond (DB) electrons, fabricated by partially hydrogenating the GaN(10${\overline{1}}$0) and ZnO(10${\overline{1}}$0) surfaces. We find that the CDW or antiferromagnetic (AFM) order has an electronic energy gain due to a band-gap opening, thereby being more stabilized compared to the metallic state. Our semilocal DFT calculation predicts that both DB wires in GaN(10${\overline{1}}$0) and ZnO(10${\overline{1}}$0) have the same CDW ground state, whereas the hybrid DFT and EX+cRPA calculations predict the AFM ground state for the former DB wire and the CDW ground state for the latter one. It is revealed that more localized Ga DB electrons in GaN(10${\overline{1}}$0) prefer the AFM order, while less localized Zn DB electrons in ZnO(10${\overline{1}}$0) the CDW formation. Our findings demonstrate that the drastically different ground states are competing in the DB wires created on the two representative compound semiconductor surfaces.

\end{abstract}

\maketitle

\section{INTRODUCTION}

Due to the confinement of electrons in low-dimensional structures, there have been many interesting quantum phases such as charge density wave (CDW)~\cite{Peierls,Gruner,Carpinelli,Yeom1}, magnetism~\cite{Erwin,Li}, and non-Fermi-liquid ground states~\cite{Tomonaga,Blumenstein}. Specifically, the CDW is usually driven by the Fermi surface nesting or the strong coupling between an electron charge modulation and a periodic lattice distortion~\cite{Peierls,Gruner}. Meanwhile, magnetism is associated with the strong electron-electron magnetic interactions~\cite{Hubbard}. These two macroscopic quantum condensates indeed represent the competing interplay between electron-phonon and electron-electron interactions, which can occur frequently in one-dimensional (1D) electron systems~\cite{Nowadnick}.

To realize 1D electron systems, the adsorption of metal atoms on semiconductor surfaces has been widely employed~\cite{Yeom1,Erwin}. For example, the In wires on Si(111)~\cite{Yeom1} and Au wires on Si(553)~\cite{Erwin} have offered unique playgrounds to search for the CDW and antiferromagnetic (AFM) orders, respectively. In contrast with such quasi-1D systems that feature atomic wires of several-atom width, a variant of hydrogen resist STM nanolithography technique, termed feedback controlled lithography~\cite{Lyding,Shen,Hersam}, has been used to generate a quasi-1D wire composed of dangling-bond (DB) electrons by selectively removing H atoms from an H-passivated Si(001) surface along one side of an Si dimer row~\cite{Hitosugi,Raza,Kepenekian,Kepenekian2,Bohloul,JYLee}. For this one-atom-wide Si DB wire, first-principles density-functional theory (DFT) calculations showed that the the Peierls-instability-driven CDW formation and the AFM spin ordering can be competing with respect to the wire length~\cite{JYLee}. Meanwhile, similarly fabricated DB wires on the H-passivated C(001) and Ge(001) surfaces were theoretically predicted to have different ground states with the AFM and CDW orders, respectively~\cite{JHLee}. It is thus most likely that the AFM or CDW ground state can be determined depending on the different degrees of localization of the 2$p$, 3$p$, and 4$p$ DB electrons in these C, Si, and Ge wires, respectively~\cite{JHLee}.

Recently, Zhao $et$ $al$.~\cite{Zhao} proposed a way to fabricate one-atom-wide metal wires on the H-passivated (10${\overline{1}}$0) surface of wurtzite semiconductors. They found that, as temperature increases, H atoms bonding to the surface Ga and Zn atoms can be selectively desorbed from the fully H-passivated GaN(10${\overline{1}}$0) and ZnO(10${\overline{1}}$0) surfaces, respectively. The resulting Ga and Zn DB wires are hereafter denoted as GaN(10${\overline{1}}$0)-1H and ZnO(10${\overline{1}}$0)-1H, respectively. Using the DFT calculations with the generalized-gradient approximation (GGA) functional of Perdew and Wang (PW)~\cite{Perdew}, Zhao $et$ $al$.~\cite{Zhao} predicted that the metallic 1${\times}$1 structure of these DB wires is unstable against Peierls distortion, leading to an insulating CDW ground state. Subsequently, more accurate schemes with the hybrid DFT and the exact-exchange plus correlation in the random-phase approximation (EX + cRPA) were employed to show that GaN(10${\overline{1}}$0)-1H has the AFM ground state rather than the CDW formation~\cite{SWKim}. Thus, it has been discussed that GaN(10${\overline{1}}$0)-1H and ZnO(10${\overline{1}}$0)-1H exhibit the competition between the AFM and CDW orders depending on the different localizations of Ga and Zn DB electrons~\cite{SWKim,Zhao2017}.
\begin{figure*}[ht]
\centering{ \includegraphics[width=16.0cm]{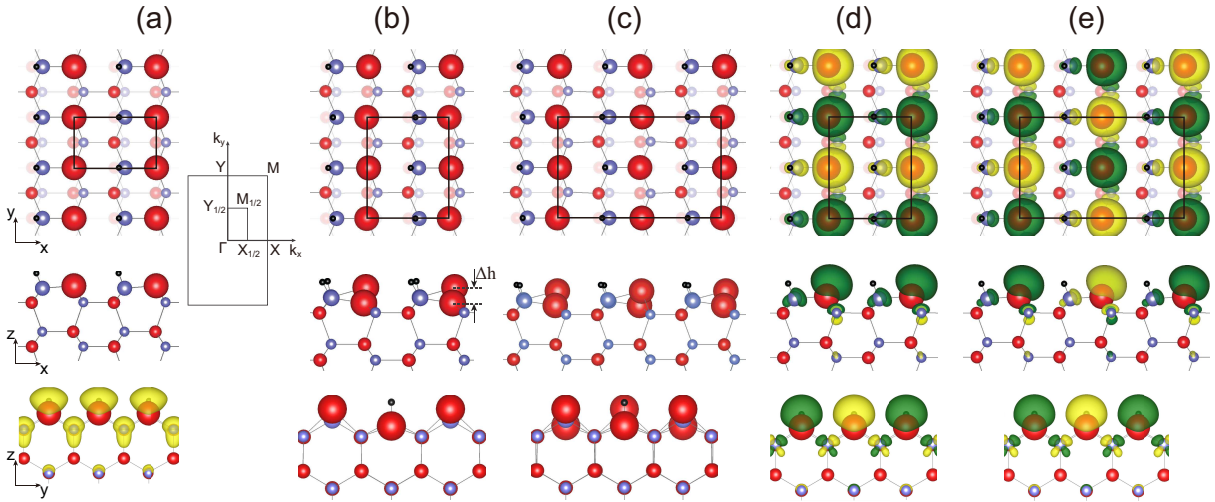} }
\caption{Top and side views of the optimized (a) NM-$p$(1${\times}$1) (b) CDW-$p$(1${\times}$2), (c) CDW-$p$(2${\times}$2), (d) AFM-$p$(1${\times}$2), and (e) AFM-$p$(2${\times}$2) structures of GaN(10${\overline{1}}$0)-1H, obtained using the PBE functional. The $x$, $y$, and $z$ axes point along the [0001], [11${\overline{2}}$0], and [10${\overline{1}}$0] directions, respectively. The red, blue, and black circles represent Ga, N and H atoms, respectively. For distinction, the surface Ga or N atoms are drawn with relatively larger circles compared to the subsurface atoms. The unit cell of each structure is indicated by the solid line. In (a), the charge density of DB electrons, obtained by integrating the occupied half-filled band, is drawn with an isosurface of 0.005 electrons/{\AA}$^3$. The surface Brillouin zones of the 1${\times}$1 and 2${\times}$2 unit cells are also drawn in (a). In (d) and (e), the spin density is drawn with an isosurface of ${\pm}$0.02 electrons/{\AA}$^3$. }
\end{figure*}

In this paper, we systematically investigate the ground states of GaN(10${\overline{1}}$0)-1H and ZnO(10${\overline{1}}$0)-1H by using the semilocal (or GGA) and hybrid DFT calculations as well as the EX + cRPA. We find that GGA predicts the CDW ground state for GaN(10${\overline{1}}$0)-1H and ZnO(10${\overline{1}}$0)-1H, in good agreement with a previous GGA calculation~\cite{Zhao}. However, both the hybrid DFT and EX + cRPA schemes predict the AFM and CDW ground states for GaN(10${\overline{1}}$0)-1H and ZnO(10${\overline{1}}$0)-1H, respectively. Our electronic-structure analysis shows that the DB electronic state in GaN(10${\overline{1}}$0)-1H is relatively more localized than that in ZnO(10${\overline{1}}$0)-1H. These different degrees of localization of the DB electrons between GaN(10${\overline{1}}$0)-1H and ZnO(10${\overline{1}}$0)-1H invoke the interplay of electron-electron and electron-phonon interactions, thereby leading to the AFM order and the CDW formation, respectively. These contrasting ground states of GaN(10${\overline{1}}$0)-1H and ZnO(10${\overline{1}}$0)-1H are anticipated to be a promising perspective in designing nanoelectronic devices on the two important and representative compound semiconductor surfaces.

\section{COMPUTATIONAL METHODS}

Our first-principles DFT calculations were performed using not only the Fritz-Haber-Institute $ab$-initio molecular simulations (FHI-aims) code~\cite{AIMS} for an accurate, all-electron description based on numeric atom-centered orbitals, but also the Vienna $ab$ $initio$ simulation package (VASP)~\cite{VASP1,VASP2} employing projector-augmented wave potentials that describe the interaction between ion cores and valence electrons~\cite{PAW}. For the VASP calculations, we used the plane wave basis with a kinetic energy cutoff of 550 eV. We treated exchange-correlation energy using the GGA functional of Perdew-Burke-Ernzerhof (PBE)~\cite{PBE} as well as the hybrid functional of Heyd-Scuseria-Ernzerhof (HSE)~\cite{HSE1,HSE2}. The HSE functional is given by
\begin{eqnarray}
E_{XC}^{\rm HSE}  =&&{\alpha}E_{X}^{\rm HF,SR}(\omega) + (1-{\alpha})E_{X}^{\rm PBE,SR}(\omega) \nonumber\\
             &&+ E_{X}^{\rm PBE,LR}(\omega) + E_{C}^{\rm PBE},
\end{eqnarray}
where the mixing factor ${\alpha}$ controls the amount of exact Fock exchange energy and the screening parameter ${\omega}$(= 0.20 {\AA}$^{-1}$) defines the separation of short range (SR) and long range (LR) for the exchange energy. It is noted that the HSE functional with ${\alpha}$ = 0 becomes identical to the PBE functional. The ${\bf k}$-space integrations in various unit-cell calculations were done equivalently with the 11${\times}$18 ${\bf k}$ points in the surface Brillouin zone of the $p$(1${\times}$1) unit cell. The GaN(10${\overline{1}}$0)-1H and ZnO(10${\overline{1}}$0)-1H surface systems were modeled by a periodic slab geometry consisting of the six atomic layers with ${\sim}$15 {\AA} of vacuum in between the slabs. Here, the bottom of the GaN(10${\overline{1}}$0)-1H slab was passivated by pseudohydrogen atoms~\cite{JZhang} with 0.75 or 1.25 $e$, while that of ZnO(10${\overline{1}}$0)-1H with 0.5 or 1.5 $e$. All atoms except the bottom two layers were allowed to relax along the calculated forces until all the residual force components were less than 0.01 eV/{\AA}. In the present EX + cRPA calculation, we considered the EX and cRPA terms using the periodic slab geometry and single-particle orbitals obtained from the PBE functional~\cite{RPA,RPA1}.

\section{RESULTS}

We first optimize the nonmagnetic (NM)-$p$(1${\times}$1) structure of GaN(10${\overline{1}}$0)-1H and ZnO(10${\overline{1}}$0)-1H using the PBE calculation. In the present study, most of the calculations were performed by using the FHI-aims code, except some analyses of

\begin{figure}[ht]
\centering{ \includegraphics[width=8.0cm]{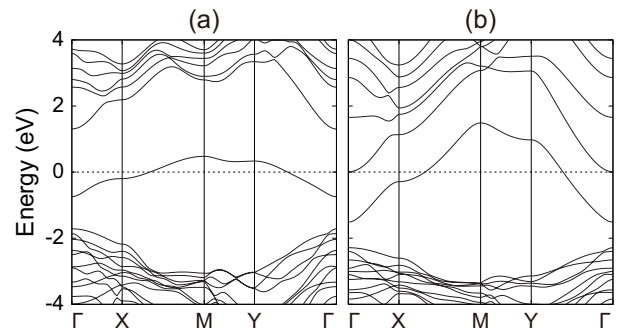} }
\caption{Calculated band structures for the NM-$p$(1${\times}$1) structures of (a) GaN(10${\overline{1}}$0)-1H and (b) ZnO(10${\overline{1}}$0)-1H. The energy zero represents the Fermi level. }
\end{figure}
\begin{figure*}[ht]
\centering{ \includegraphics[width=16.0cm]{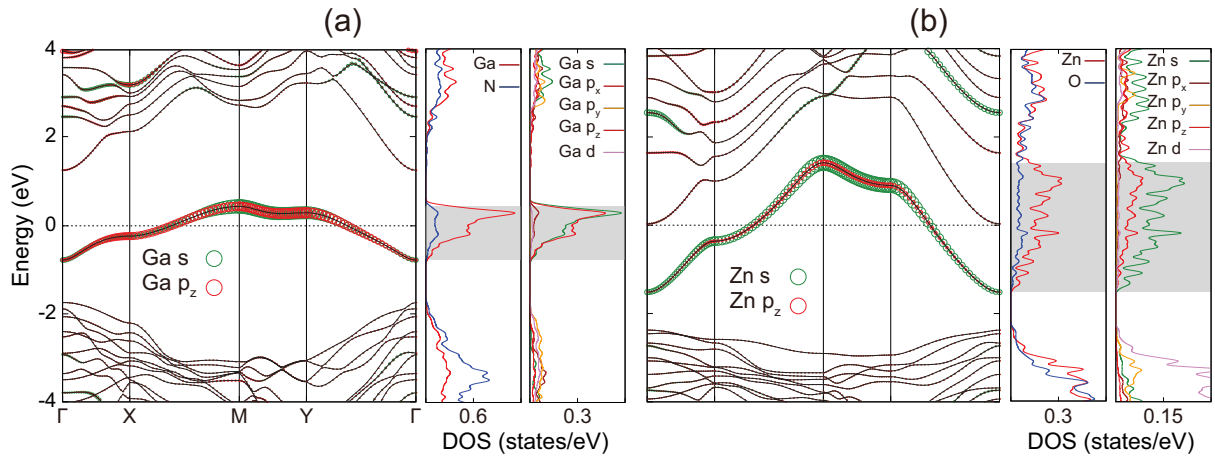} }
\caption{Calculated band structures for the NM-$p$(1${\times}$1) structures of (a) GaN(10${\overline{1}}$0)-1H and (b) ZnO(10${\overline{1}}$0)-1H, obtained using the VASP code. The bands projected onto the $s$ and $p_{\rm z}$ orbitals of the Ga (Zn) surface atoms are also displayed with circles whose radii are proportional to their weights, together with the PDOS projected onto the corresponding Ga and neighboring N (or Zn and O) orbitals. }
\end{figure*}

electronic structure using the VASP. Figure 1(a) shows the optimized NM-$p$(1${\times}$1) structure of GaN(10${\overline{1}}$0)-1H. We find that the Ga (Zn) surface atoms containing DB electrons are separated by 3.228 (3.282) and 5.262 (5.297) {\AA} along the $y$ and $x$ directions parallel and perpendicular to the wires, respectively. The corresponding band structures of GaN(10${\overline{1}}$0)-1H and ZnO(10${\overline{1}}$0)-1H are displayed in Figs. 2(a) and (b), respectively. It is seen that GaN(10${\overline{1}}$0)-1H and ZnO(10${\overline{1}}$0)-1H have a half-filled band with the bandwidth of 1.22 and 3.00 eV, respectively. As shown in Fig. 1(a), the charge character of the half-filled band, integrated downward from the Fermi level $E_{\rm F}$, represents the DB electrons along the wires [see Fig. 1(a)]. To figure out the detailed features of the half-filled band, we use the VASP code to calculate the band projection onto the Ga (or Zn) surface atoms as well as the partial density of states (PDOS) projected onto the corresponding Ga and neighboring N (or Zn and O) orbitals. The results for GaN(10${\overline{1}}$0)-1H are given in Fig. 3(a), while those for ZnO(10${\overline{1}}$0)-1H in Fig. 3(b), respectively. For GaN(10${\overline{1}}$0)-1H, the PDOS of the Ga $s$, $p_x$, $p_y$, and $p_z$ orbitals near $E_{\rm F}$ gives a ratio of 48\%: 5\%: 0\%: 47\% [see Fig. 3(a)], whereas for ZnO(10${\overline{1}}$0)-1H the corresponding PDOS of the Zn $s$, $p_x$, $p_y$, and $p_z$ orbitals is 71\%: 3\%: 1\%: 25\% [Fig. 3(b)]. Therefore, the half-filled band of GaN(10${\overline{1}}$0)-1H has a nearly equal mixing character of 4$s$ and 4$p_z$ orbitals [see Fig. 3(a)], while that of ZnO(10${\overline{1}}$0)-1H has a much greater 4$s$ orbital character than 4$p_z$ orbital [Fig. 3(b)]. These contrasting orbital characters of the half-filled band between GaN(10${\overline{1}}$0)-1H and ZnO(10${\overline{1}}$0)-1H may reflect their different bandwidths: i.e. the former DB state is more localized than the latter one, indicating the different degrees of localization of the DB electrons in the two systems.

\begin{table}[ht]
\caption{Calculated total energies [in meV per $p$(1${\times}$1) unit cell] of different unit-cell structures relative to NM-$p$(1${\times}$1), in comparison with previous calculations~\cite{Zhao,Zhao2017}. In the present HSE calculation, we use the optimal mixing factor ${\alpha}_{\rm opt}$ = 0.32 and 0.20 for GaN(10${\overline{1}}$0)-1H and ZnO(10${\overline{1}}$0)-1H, respectively. Meanwhile, the HSE calculation of Zhao $et$ $al$.~\cite{Zhao2017} used the same value of ${\alpha}$ = 0.25 for GaN(10${\overline{1}}$0)-1H and ZnO(10${\overline{1}}$0)-1H. }
\begin{ruledtabular}
\begin{tabular}{llcccc}
                             &                         & PBE     & HSE      & GGA-PW~\cite{Zhao} & HSE~\cite{Zhao2017}  \\ \hline
GaN(10${\overline{1}}$0)-1H  & CDW-$p$(1${\times}$2)   & $-$133  & $-$225   & $-$128             & $-$          \\
                             & AFM-$p$(1${\times}$2)   & $-$81   & $-$272   & $-$                & $-$          \\
                             & CDW-$p$(2${\times}$2)   & $-$147  & $-$242   & $-$143             & $-$218  \\
                             & AFM-$p$(2${\times}$2)   & $-$88   & $-$275   & $-$                & $-$233  \\
                             & FM-$p$(1${\times}$1)    & $-$33   & $-$231   & $-$                & $-$          \\ \hline
ZnO(10${\overline{1}}$0)-1H  & CDW-$p$(1${\times}$2)   & $-$     & $-$1.1   & $-$                & $-$          \\
                             & AFM-$p$(1${\times}$2)   & $-$     & $-$0.8   & $-$                & $-$          \\
                             & CDW-$p$(2${\times}$2)   & $-$10.9 & $-$25.8  & $-$25              & $-$60     \\
                             & AFM-$p$(2${\times}$2)   & 0.1     & $-$21.3  & $-$                & $-$50     \\
\end{tabular}
\end{ruledtabular}
\end{table}
In order to obtain the more stable structures compared to the NM-$p$(1${\times}$1) structure with the half-filled band, we optimize the two competing structures including the CDW and AFM orders using the PBE calculation. These two orders driven by electron-phonon and electron-electron interactions usually have a band-gap opening with a doubled periodicity along the wire direction~\cite{JYLee,JHLee}. The calculated total energies of different unit-cell structures relative to the NM-$p$(1${\times}$1) structure are given in Table I. For GaN(10${\overline{1}}$0)-1H, the CDW-$p$(1${\times}$2) [see Fig. 1(b)] and AFM-$p$(1${\times}$2) [Fig. 1(d)] structures are found to be more stable than the NM-$p$(1${\times}$1) structure by 133 and 81 meV per $p$(1${\times}$1) unit cell. However, for ZnO(10${\overline{1}}$0)-1H, we were unable to find the stable CDW-$p$(1${\times}$2) and AFM-$p$(1${\times}$2) structures. This absence of the CDW-$p$(1${\times}$2) or AFM-$p$(1${\times}$2) structure in ZnO(10${\overline{1}}$0)-1H may be ascribed to the self-interaction error (SIE)~\cite{Cohen} inherent in the PBE functional, which does not properly describe the CDW-$p$(1${\times}$2) or AFM-$p$(1${\times}$2) structure but artificially stabilizes the NM-$p$(1${\times}$1) structure due to its reduced self-repulsion caused by an over-delocalization of the half-filled state~\cite{SIE}. We will show below that this SIE in ZnO(10${\overline{1}}$0)-1H can be cured by the HSE functional to obtain the stable CDW-$p$(1${\times}$2) and AFM-$p$(1${\times}$2) structures. It is noteworthy that the half-filled bands in the NM-$p$(1${\times}$1) structures of GaN(10${\overline{1}}$0)-1H and ZnO(10${\overline{1}}$0)-1H are dispersive along not only the $\overline{{\Gamma}Y}$ line but also the $\overline{{\Gamma}X}$ line, as shown in Figs. 2(a) and 2(b). This indicates that Ga and Zn DB wires have some interactions along the $x$ direction perpendicular to the wires. Such inter-wire interactions are found to give an additional energy lowering with a doubled periodicity along the $x$ direction. Our PBE calculation for GaN(10${\overline{1}}$0)-1H shows that the CDW-$p$(2${\times}$2) [see Fig. 1(c)] and AFM-$p$(2${\times}$2) [Fig. 1(e)] structures are further thermodynamically favored over the NM-$p$(1${\times}$1) structure by 147 and 88 meV per $p$(1${\times}$1) unit cell. Here, the geometry of AFM-$p$(2${\times}$2) changes relative to NM-$p$(1${\times}$1) by less than 0.01 {\AA}. Therefore, the AFM-$p$(2${\times}$2) phase has a symmetry breaking due to the spin ordering. For ZnO(10${\overline{1}}$0)-1H, the CDW-$p$(2${\times}$2) structure becomes more stable than the NM-$p$(1${\times}$1) structure by 10.9 meV per $p$(1${\times}$1) unit cell, while the AFM-$p$(2${\times}$2) structure is almost degenerate with the NM-$p$(1${\times}$1) structure. Thus, our PBE calculations for GaN(10${\overline{1}}$0)-1H and ZnO(10${\overline{1}}$0)-1H predict the CDW-$p$(2${\times}$2) ground state, consistent with a previous GGA-PW calculation~\cite{Zhao} (see Table I).

\begin{figure}[ht]
\centering{ \includegraphics[width=8.0cm]{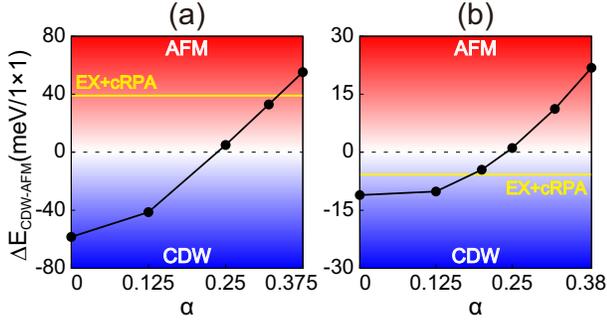} }
\caption{Calculated total energy difference ${\Delta}E_{\rm CDW-AFM}$ between CDW-$p$(2${\times}$2) and AFM-$p$(2${\times}$2) for (a) GaN(10${\overline{1}}$0)-1H and (b) ZnO(10${\overline{1}}$0)-1H. In (a) and (b), the EX+cRPA results are indicated by yellow lines.}
\end{figure}

To cure the above-mentioned SIE in the PBE results, we perform the HSE calculations for GaN(10${\overline{1}}$0)-1H and ZnO(10${\overline{1}}$0)-1H as a function of ${\alpha}$. The calculated total energy difference ${\Delta}E_{\rm CDW-AFM}$ between the CDW-$p$(2${\times}$2) and AFM-$p$(2${\times}$2) structures for GaN(10${\overline{1}}$0)-1H and ZnO(10${\overline{1}}$0)-1H is displayed as a function of ${\alpha}$ in Fig. 4(a) and 4(b), respectively. We find that, as ${\alpha}$ increases, ${\Delta}E_{\rm CDW-AFM}$ for GaN(10${\overline{1}}$0)-1H and ZnO(10${\overline{1}}$0)-1H monotonically increases, leading to the AFM $p$(2${\times}$2) ground state above a critical value of ${\alpha}$ ${\approx}$ 0.25. In order to find the optimal value ${\alpha}_{\rm opt}$, we perform more rigorous EX + cRPA calculations for the CDW-$p$(2${\times}$2) and AFM-$p$(2${\times}$2) structures. The EX + cRPA results for ${\Delta}E_{\rm CDW-AFM}$ of GaN(10${\overline{1}}$0)-1H and ZnO(10${\overline{1}}$0)-1H are displayed in Fig. 4(a) and 4(b), respectively. We find that ${\Delta}E_{\rm CDW-AFM}$ is 39 ($-$6) meV per $p$(1${\times}$1) unit cell for GaN(10${\overline{1}}$0)-1H (ZnO(10${\overline{1}}$0)-1H), which is close to the HSE result with ${\alpha}_{\rm opt}$ ${\approx}$ 0.32 (0.20). Thus, our EX + cRPA calculation predicts the AFM-$p$(2${\times}$2) and CDW-$p$(2${\times}$2) ground states for GaN(10${\overline{1}}$0)-1H and ZnO(10${\overline{1}}$0)-1H, respectively.

In Table I, we list the energetics of various states of GaN(10${\overline{1}}$0)-1H and ZnO(10${\overline{1}}$0)-1H, obtained using the HSE functional with ${\alpha}_{\rm opt}$. For GaN(10${\overline{1}}$0)-1H, HSE predicts that the CDW-$p$(1${\times}$2), CDW-$p$(2${\times}$2), AFM-$p$(1${\times}$2), and AFM-$p$(2${\times}$2) structures are more stable than the NM-$p$(1${\times}$1) structure by 225, 242, 272, and 275 meV per $p$(1${\times}$1) unit cell, which are larger than the corresponding ones obtained using PBE (see Table I). Especially, HSE much enhances the stabilization of the AFM order relative to the NM-$p$(1${\times}$1) structure, compared to PBE. Therefore, the HSE results show that AFM-$p$(1${\times}$2) is preferred over CDW-$p$(1${\times}$2) by 47 meV, slightly larger than that (33 meV) between AFM-$p$(2${\times}$2) and CDW-$p$(2${\times}$2). Meanwhile, for ZnO(10${\overline{1}}$0)-1H, HSE predicts that the CDW-$p$(2${\times}$2) and AFM-$p$(2${\times}$2) structures are more stable than the NM-$p$(1${\times}$1) structure by 25.8 and 21.3 meV per $p$(1${\times}$1) unit cell, but CDW-$p$(1${\times}$2) and AFM-$p$(1${\times}$2) are nearly degenerate with NM-$p$(1${\times}$1). Therefore, contrasting with the AFM-$p$(2${\times}$2) ground state in GaN(10${\overline{1}}$0)-1H, the CDW-$p$(2${\times}$2) structure is found to be the most stable in ZnO(10${\overline{1}}$0)-1H.

\begin{figure}[ht]
\centering{ \includegraphics[width=8.0cm]{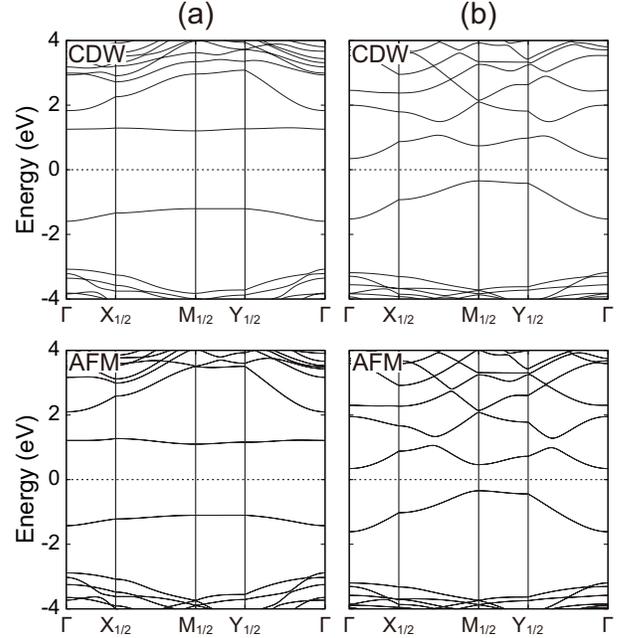} }
\caption{Calculated HSE band structures for CDW-$p$(2${\times}$2) and AFM-$p$(2${\times}$2) of (a) GaN(10${\overline{1}}$0)-1H and (b) ZnO(10${\overline{1}}$0)-1H. The energy zero represents the Fermi level.}
\end{figure}

Figures 5(a) and 5(b) show the HSE band structures for the CDW-$p$(2${\times}$2) and AFM-$p$(2${\times}$2) structures of GaN(10${\overline{1}}$0)-1H and ZnO(10${\overline{1}}$0)-1H, respectively. For GaN(10${\overline{1}}$0)-1H, the band gap of CDW-$p$(2${\times}$2) is found to be 2.41 eV, which is slightly larger than that (2.20 eV) of AFM-$p$(2${\times}$2). This indicates that the electronic energy gain due to a band-gap opening is relatively larger in CDW-$p$(2${\times}$2) compared to AFM-$p$(2${\times}$2). It is, however, noted that the surface Ga atoms in CDW-$p$(2${\times}$2) are displaced up and down alternatively with a height difference ${\Delta}{\rm h}$ of 0.87 {\AA} [see Fig. 1(b)], whereas such a lattice distortion is absent in the AFM-$p$(2${\times}$2) structure, i.e., the surface Ga atoms have an equal height [see Fig. 1(e)]. The large Peierls-like distortion of CDW-$p$(2${\times}$2) may give an elastic energy cost of the strains, thereby leading to the AFM-$p$(2${\times}$2) ground state in GaN(10${\overline{1}}$0)-1H. This preference for the AFM order over the CDW formation is likely to be attributed to a large localization of the DB electrons in GaN(10${\overline{1}}$0)-1H, as discussed earlier. Indeed, the band widths for the Ga DB electrons in CDW-$p$(2${\times}$2) and AFM-$p$(2${\times}$2) are very narrow with less than ${\sim}$0.36 eV [see Fig. 5(a)]. On the other hand, as shown in Fig. 5(b), the corresponding band widths for the Zn DB electrons are broad with ${\sim}$1.22 eV. Interestingly, the band gaps of CDW-$p$(2${\times}$2) and AFM-$p$(2${\times}$2) in ZnO(10${\overline{1}}$0)-1H are also nearly equal to each other as ${\sim}$0.69 eV [see Fig. 5(b)], implying a similar electronic energy gain. However, some delocalized character of the Zn DB state can reduce the Peierls distortion of CDW-$p$(2${\times}$2): i.e., the buckling of the surface Zn atoms is calculated to be 0.59 {\AA}, much smaller than the corresponding one (0.87 {\AA}) of the surface Ga atoms in GaN(10${\overline{1}}$0)-1H. It is thus  likely that such delocalized Zn DB electrons prefer to have electron-phonon coupling, giving rise to the CDW-$p$(2${\times}$2) ground state in ZnO(10${\overline{1}}$0)-1H.

\begin{figure}[ht]
\centering{ \includegraphics[width=8.0cm]{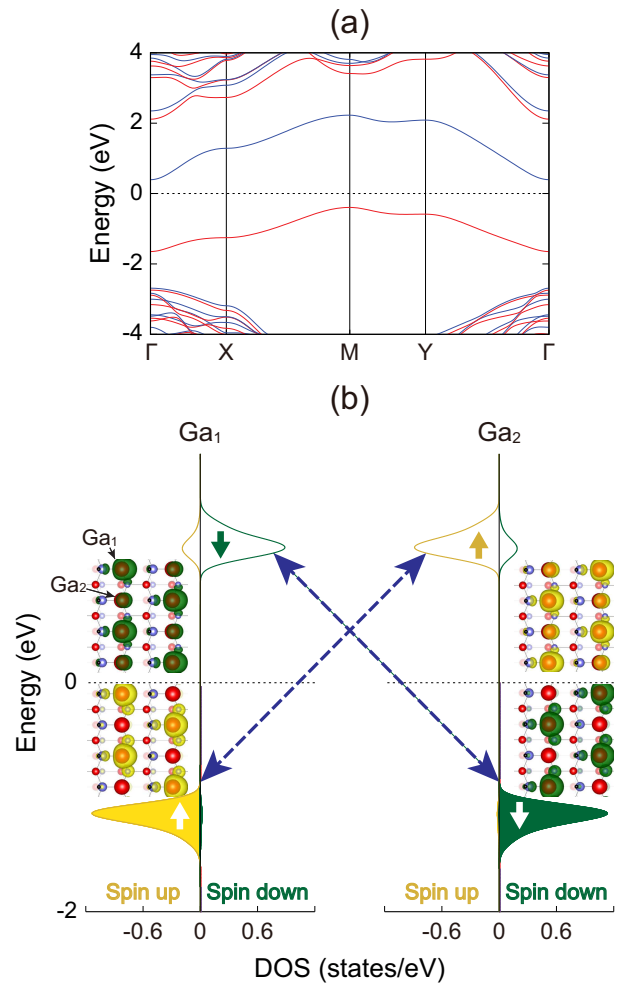} }
\caption{(a) Calculated band structure of FM-$p$(1${\times}$1). The spin-polarized local DOS projected onto two neighboring Ga surface atoms (Ga$_1$ and Ga$_2$) in the AFM-$p$(2${\times}$2) structure of GaN(10${\overline{1}}$0)-1H is displayed in (b). The charge characters of the spin-up and spin-down states for the highest occupied and the lowest unoccupied bands are taken at the $\Gamma$ point with an isosurface of 0.01 electrons/{\AA}$^3$. The energy zero represents the Fermi level.}
\end{figure}

It is noteworthy that, according to the HSE calculation for GaN(10${\overline{1}}$0)-1H, the CDW-$p$(1${\times}$2) and AFM-$p$(1${\times}$2) structures are much more stabilized relative to the NM-$p$(1${\times}$1) structure by 225 and 272 meV per $p$(1${\times}$1) unit cell, respectively, while the CDW-$p$(2${\times}$2) and AFM-$p$(2${\times}$2) structures give rise to small additional energy gains of 17 and 3 meV per $p$(1${\times}$1) unit cell (see Table I). On the other hand, for ZnO(10${\overline{1}}$0)-1H, the energy gains for CDW-$p$(1${\times}$2) and AFM-$p$(1${\times}$2) relative to NM-$p$(1${\times}$1) are negligible compared to those of CDW-$p$(2${\times}$2) and AFM-$p$(2${\times}$2). These drastic different aspects between GaN(10${\overline{1}}$0)-1H and ZnO(10${\overline{1}}$0)-1H imply that GaN(10${\overline{1}}$0)-1H has a quasi-1D feature with a strong intra-wire interaction, whereas ZnO(10${\overline{1}}$0)-1H a quasi-2D feature with a rather strong inter-wire interaction. Here, we note that, although the concept of a Fermi surface nesting-induced CDW originated from the Peierls idea of electronic instabilities in purely 1D metals~\cite{Peierls}, it has been extended to include structural phase transitions driven by the so-called $q$-dependent electron-phonon coupling, thereby leading to no meaningful distinction between a CDW and a structural phase transition at the surface~\cite{johan}.

Figures 1(d) and 1(e) show the PBE spin densities for the AFM-$p$(1${\times}$2) and AFM-$p$(2${\times}$2) structures of GaN(10${\overline{1}}$0)-1H, respectively. It is seen that the spin moments are highly localized at the Ga surface atoms. Using a Mulliken population analysis~\cite{mulliken}, we obtain the PBE (HSE) spin moment of each Ga surface atom for the AFM-$p$(1${\times}$2) and AFM-$p$(2${\times}$2) structures as 0.67 (0.79) and 0.66 (0.78) ${\mu}_{\rm B}$, respectively. It is noticeable that the spin moments of two neighboring Ga surface atoms are antiferromagnetically coupled along the $y$ direction through an interconnecting N subsurface atom [see Figs. 1(d) and 1(e)]. The resulting symmetry reduction at the surface possibly induces a magnetic anisotropy, which in turn lifts the Mermin-Wagner restriction~\cite{mermin} that prohibits magnetic order in the 2D isotropic Heisenberg model at finite temperatures. Although the quantitative estimation of the magnetic anisotropy in the GaN(10${\overline{1}}$0)-1H system is out of the scope of the present study, the AFM order is likely to be stabilized at finite temperatures. Indeed, a variety of 2D layered magnetic compounds have recently been studied to demonstrate that their magnetic properties can be retained down to monolayer thickness~\cite{gong,huang}.

To examine the exchange interaction energy between the magnetic moments of Ga surface atoms in GaN(10${\overline{1}}$0)-1H, we perform the HSE (with ${\alpha}_{\rm opt}$) calculation for the ferromagnetic (FM)-$p$(1${\times}$1) structure. We find that FM-$p$(1${\times}$1) is less stable than AFM-$p$(2${\times}$2) by 44 meV per $p$(1${\times}$1) unit cell (see Table I). In Fig. 6(a), the band structure of FM-$p$(1${\times}$1) shows a band gap of 0.79 eV, much smaller than that (2.20 eV) of AFM-$p$(2${\times}$2). This indicates that the AFM-$p$(2${\times}$2) structure gives a relatively larger electronic energy gain, compared to the FM one. Using the Heisenberg Hamiltonian, the exchange coupling constant $J$ between neighboring magnetic moments is estimated to be as large as ${\sim}$22 meV, indicating a strong AFM exchange coupling. Figure 6(b) shows the spin-polarized local DOS projected onto two neighboring Ga surface atoms (Ga$_1$ and Ga$_2$) in the AFM-$p$(2${\times}$2) structure of GaN(10${\overline{1}}$0)-1H as well as their spin characters. It is seen that the occupied spin-up and spin-down states are localized at Ga$_1$ and Ga$_2$, while the unoccupied spin-up and spin-down states at Ga$_2$ and Ga$_1$, respectively. Considering the fact that the hybridization occurs between electronic states with the same spin direction, the occupied and unoccupied spin-up or spin-down states can hybridize with each other. This exchange interaction mediated by the occupied and unoccupied electronic states is characterized as a superexchange mechanism~\cite{Goodenough,Kanamori,Sato}.

\section{SUMMARY}

Using the PBE, HSE, and EX + cRPA calculations, we have systematically investigated the competing CDW and AFM phases of GaN(10${\overline{1}}$0)-1H and ZnO(10${\overline{1}}$0)-1H. Our PBE calculation predicts the same CDW-$p$(2${\times}$2) ground state for GaN(10${\overline{1}}$0)-1H and ZnO(10${\overline{1}}$0)-1H. However, both the HSE and EX + cRPA calculations predict the AFM-$p$(2${\times}$2) and CDW-$p$(2${\times}$2) ground states for GaN(10${\overline{1}}$0)-1H and ZnO(10${\overline{1}}$0)-1H, respectively. It is revealed that the DB state in GaN(10${\overline{1}}$0)-1H is relatively more localized than that in ZnO(10${\overline{1}}$0)-1H. These different degrees of localization of the DB electrons between GaN(10${\overline{1}}$0)-1H and ZnO(10${\overline{1}}$0)-1H are likely to cause the interplay of electron-electron and electron-phonon interactions, leading to the drastically different AFM and CDW ground states in the two systems. The future experimental works are stimulated to validate these theoretical predictions.

\vspace{0.4cm}
\centerline{\bf ACKNOWLEDGEMENTS}
\vspace{0.4cm}

This work was supported by the National Research Foundation of Korea grant funded by the Korean government (Nos. 2015M3D1A1070639 and 2015R1A2A2A01003248), KISTI supercomputing center through the strategic support program for the supercomputing application research (KSC-2017-C3-0041). S.W.K. acknowledges support from POSCO TJ Park Foundation and Y.G.K. acknowledges support from Hyundai motor Chung Mong-Koo Foundation.

\noindent $^{*}$ Corresponding authors: chojh@hanyang.ac.kr

\end{document}